\documentclass[preprint,showpacs,superscriptaddress,nofootinbib]{revtex4}
\usepackage[dvips]{graphicx}
\usepackage{amsmath,amssymb}

\newcommand{\MNS}{{\text{MNS}}}

\newcommand{\eV}{{\text{eV}}}

\newcommand{\GeV}{{\text{GeV}}}
\newcommand{\TeV}{{\text{TeV}}}
\newcommand{\BR}{\text{BR}}

\newcommand{\meg}{\mu \to e \gamma}

\newcommand{\pell}{{\ell^\prime}}

\newcommand{\Uee}{(U_\MNS)_{e 1}}
\newcommand{\Uem}{(U_\MNS)_{e 2}}
\newcommand{\Uet}{(U_\MNS)_{e 3}}
\newcommand{\Ume}{(U_\MNS)_{\mu 1}}
\newcommand{\Umm}{(U_\MNS)_{\mu 2}}
\newcommand{\Umt}{(U_\MNS)_{\mu 3}}
\newcommand{\Ute}{(U_\MNS)_{\tau 1}}
\newcommand{\Utm}{(U_\MNS)_{\tau 2}}
\newcommand{\Utt}{(U_\MNS)_{\tau 3}}

\newcommand{\themodel}{{1LDNM}}

\begin{document}

\preprint{UT-HET 054}

\title{
 Neutrino Masses
from Loop-Induced Dirac Yukawa Couplings
}

\author{Shinya Kanemura}
\email{kanemu@sci.u-toyama.ac.jp}
\affiliation{
Department of Physics,
University of Toyama, Toyama 930-8555, Japan
}
\author{Takehiro Nabeshima}
\email{nabe@jodo.sci.u-toyama.ac.jp}
\affiliation{
Department of Physics,
University of Toyama, Toyama 930-8555, Japan
}
\author{Hiroaki Sugiyama}
\email{hiroaki@fc.ritsumei.ac.jp}
\affiliation{
Department of Physics,
Ritsumeikan University, Kusatsu, Shiga 525-8577, Japan
}


\begin{abstract}
 We consider a possibility
to naturally explain tiny neutrino masses
without the lepton number violation.
 We study a simple model
with $SU(2)_L$ singlet charged scalars ($s_1^\pm, s_2^\pm$)
as well as singlet right-handed neutrino ($\nu_R$).
 Yukawa interactions for Dirac neutrinos,
which are forbidden at the tree level
by a softly-broken $Z_2$ symmetry,
are induced at the one-loop level
via the soft-breaking term in the scalar potential.
 Consequently neutrinos obtain small Dirac masses
after the electroweak symmetry breaking.
 It is found that constrains from neutrino oscillation measurements
and lepton flavor violation search results (especially for $\meg$)
can be satisfied.
 We study the decay pattern of the singlet charged scalars,
which could be tested at the LHC and the ILC\@.
 We discuss possible extensions also,
e.g.\ to introduce dark matter candidate.
\end{abstract}


\maketitle

\section{introduction}
\label{sec:intro}

 The neutrino oscillation data
provide evidence that
neutrinos have tiny masses~\cite{solar,Wendell:2010md,acc,Apollonio:2002gd,Gando:2010aa},
which can be understood as a clear signature for physics
beyond the standard model~(SM).
 The simplest way to obtain neutrino masses
may be to introduce $SU(2)_L$-singlet right-handed neutrinos,
$\nu_R^i$ $(i=1\text{-}3)$,
which have Yukawa interaction with the SM Higgs boson.
 Then Dirac masses for neutrinos
are generated after the electroweak symmetry breaking.
 However, in this naive mechanism,
the Yukawa coupling constants for neutrinos
have to be unnaturally small ($\lesssim 10^{-11}$)
in comparison with those for the other fermions.
 The most familiar idea to solve the problem
would be the seesaw mechanism~\cite{seesaw}
by introducing Majorana mass terms for $\nu_R^i$.
 Taking the Majorana masses much larger than
vacuum expectation value of the SM Higgs boson,
very light Majorana neutrinos are obtained
without excessively small Yukawa coupling constants.

 Another interesting possibility
to avoid tiny Yukawa coupling constants
is the radiative generation of Majorana masses for $\nu_L$
without Dirac mass terms.
 The original model was proposed by Zee~\cite{Zee:1980ai},
in which Majorana neutrino masses
are obtained at the one-loop level
by the dynamics of the extended Higgs sector%
\footnote
{
 If leptons couple with only one of two doublet scalar fields
in the Zee model in order to eliminate
the flavor changing neutral current interaction,
the model cannot satisfy neutrino oscillation data~\cite{He:2003ih}.
}.
 There have been several variant models
in this direction~\cite{ZB,Krauss:2002px,Ma:2006km,Aoki:2008av}%
\footnote
{
 In ref.~\cite{Cheung:2004xm},
the second singlet fermion is added
to the model in ref.~\cite{Krauss:2002px}
in order to satisfy neutrino oscillation data.
}.
 Some of them~\cite{Krauss:2002px,Ma:2006km,Aoki:2008av}
include dark matter candidates
by imposing an unbroken $Z_2$ symmetry
which forbids the Dirac masses for neutrinos.
 In those models (seesaw and radiative ones),
neutrinos are regarded as Majorana particles
whose mass terms cause lepton number violating phenomena
such as the neutrinoless double beta decay.
 Lepton number violation~(LNV), however,
has not yet been confirmed by experiments.
 Thus
it is valuable to investigate a possibility
that the tiny neutrino masses are generated
in theories where the lepton number is conserved
and Yukawa coupling constants are not extremely small.

 Radiative generation of masses for Dirac neutrinos
would be an interesting possibility.
 There were several studies in past for such a scenario
in various frameworks such as
the left-right symmetry%
~\cite{Mohapatra:1987hh,1loopLR},
supersymmetry~(SUSY)~\cite{1loopSUSY},
and extended models
within the SM gauge group~\cite{Nasri:2001ax,Gu:2007ug}
(See also ref.~\cite{Babu:1989fg}).
 The simplest model
seems to be the one in ref.~\cite{Nasri:2001ax},
where Dirac neutrino masses are generated at the one-loop level
by introducing two $SU(2)_L$-singlet charged scalar fields
($s_1^\pm, s_2^\pm$) as well as $\nu_R^i$.
 In this letter,
we show the one-loop Dirac neutrino model~(\themodel)
is compatible with neutrino oscillation data
although this was overlooked in \cite{Nasri:2001ax}.
 We can find parameter sets
which satisfy also constraints
from searches for lepton flavor violation.
 We discuss the collider phenomenology
of charged scalars under these parameter sets.
 Their decay pattern into leptons
can be a characteristic feature of the \themodel,
by which the model could be tested at the LHC
and the International Linear Collider~(ILC).

 We also discuss
some extensions of the model briefly;
accommodating dark matter candidates,
case with Majorana masses for $\nu_R$,
and so on.

\section{the model}
\label{sec:model}

\begin{table}[t]
\begin{center}
\begin{tabular}{c||c|c|c||c|c|c}
 {}
 & \
 $L_\ell =
 \begin{pmatrix}
  \nu_L^\ell\\
  \ell_L
 \end{pmatrix}$ \
 & \ $\ell_R$ \
 & \
 $\Phi =
 \begin{pmatrix}
  \phi^+\\
  \phi^0
 \end{pmatrix}$ \
 & \ $\nu_R^i$ \
 & \ $s_1^+$ \
 & \ $s_2^+$ \
\\\hline\hline
 \ $SU(2)_L$ \
 & \ ${\bf \underline{2}}$ \
 & \ ${\bf \underline{1}}$ \
 & \ ${\bf \underline{2}}$ \
 & \ ${\bf \underline{1}}$ \
 & \ ${\bf \underline{1}}$ \
 & \ ${\bf \underline{1}}$ \
\\\hline
 \ $Y$ \
 & \ $1/2$ \
 & \ $-1$ \
 & \ $1/2$ \
 & \ $0$ \
 & \ $1$ \
 & \ $1$ \
\\\hline
 \ softly-broken $Z_2$ \
 & \ $+$ \
 & \ $+$ \
 & \ $+$ \
 & \ $-$ \
 & \ $+$ \
 & \ $-$ \
\\\hline
 \ lepton number \
 & \ $1$ \
 & \ $1$ \
 & \ $0$ \
 & \ $1$ \
 & \ $-2$ \
 & \ $-2$ \
\end{tabular}
\end{center}
\caption{
 Particle contents of the \themodel.
 Here
$L_\ell$, $\ell_R$, and $\Phi$
are the $SU(2)_L$-doublet fields
of left-handed leptons,
the right-handed charged leptons,
and the $SU(2)_L$-doublet scalar field in the SM,
respectively.
 Three column from the right show particles
added to the SM\@.
}
\label{tab:particle}
\end{table}

 Particle contents of the \themodel\ are
listed in Table~\ref{tab:particle}.
 Three $SU(2)_L$-singlet neutral fermions
$\nu_R^i$ ($i = 1\text{-}3$)
are introduced
such that Dirac masses for neutrinos exist.
 A softly-broken $Z_2$ symmetry is imposed in the model
so that Dirac masses can be forbidden at the tree level,
where $\nu_R^i$ are assigned to be $Z_2$-odd.
 Dirac neutrino masses are generated at the one-loop level
by utilizing $SU(2)_L$-singlet charged scalars,
$s_1^+$ and $s_2^+$,
where $s_2^+$ is taken as a $Z_2$-odd particle
which can couple with $\nu_R^i$.
 The Yukawa interactions, the Higgs potential,
and the Dirac neutrino mass generation in this model
are presented below in order.

 The Yukawa interactions for leptons are given by
\begin{eqnarray}
{\mathcal L}_{\text{Yukawa}}
&=&
 y_\ell\,
 \overline{L_\ell}\,\Phi\,\ell_R
 +
 f_{\ell\pell}\,
 \overline{L_\ell^c}\,i\sigma_2\,L_{\pell}\,s_1^+
 +
 h_{\ell i}\,
 \overline{(\ell_R)^c}\,\nu_R^i\,s_2^+
 +
 \text{h.c.} ,
\end{eqnarray}
where $\Phi$ is the SM Higgs doublet field.
The superscript $c$ denotes the charge conjugation
and $\sigma_i$ ($i=1\text{-}3$) are the Pauli matrices.
 We take the basis where the Yukawa coupling matrix for charged leptons
has been diagonalized as $y_\ell^{}$.
 Notice that the matrix $f$ is antisymmetric,
while the matrix $h$ takes somehow an arbitrary form.
 Although $h_{\ell i}$ and $f_{\ell\pell}$
are basically complex numbers,
$f_{\ell\pell}$ can be taken to be real numbers
by using rephasing of three $L_\ell$
(and $\ell_R$ to keep $y_\ell^{}$ real)
without loss of generality.
 Furthermore
we can take the basis where $\nu_R^i$ are mass eigenstates
(of real positive mass eigenvalues).
 Then
neutrino oscillation data
give relations between
the elements of $f_{\ell\pell}$ and $h_{\ell i}$
as shown later.

 The Higgs potential is written as
\begin{eqnarray}
V
&=&
 - \mu^2\, \Phi^\dagger \Phi
 + \lambda (\Phi^\dagger \Phi)^2
 + \mu_1^2\, |s_1^+|^2
 + \mu_2^2\, |s_2^+|^2
 + \left\{ \mu_3^2\, s_1^+ s_2^- + \text{h.c.} \right\}
\nonumber\\
&&\hspace*{5mm}
{}
 + \lambda_1\, |s_1^+|^4
 + \lambda_2\, |s_2^+|^4
 + \left\{ \lambda_3\, (s_1^+ s_2^-)^2 + \text{h.c.} \right\}
 + \lambda_4\, |s_1^+|^2 |s_2^+|^2
\nonumber\\
&&\hspace*{5mm}
{}
 + \lambda_5\, (\Phi^\dagger \Phi) |s_1^+|^2
 + \lambda_6\, (\Phi^\dagger \Phi) |s_2^+|^2 ,
\end{eqnarray}
where $\mu^2 > 0$.
 Although $\mu_3^2$ and $\lambda_3$ can be complex parameters,
they become real by rephasing $s_1^+$ and $s_2^+$.
 Thus
there is no complex parameter in the Higgs potential.
 Notice that
$\mu_3^2$ is the soft-breaking parameter
for the $Z_2$ symmetry we imposed.
 The quartic coupling constants should satisfy
the following conditions
in order to avoid that
the potential is unbounded from below;
\begin{eqnarray}
&&
\lambda > 0 , \quad
\lambda_1 > 0 , \quad
\lambda_2 > 0 ,
\\
&&
\omega_1
\equiv
 2\lambda_1
 + \lambda_5 \sqrt{\frac{\lambda_1}{\lambda}} > 0 , \quad
2\sqrt{\omega_1 \lambda_2}
+ 2\lambda_3
+ \lambda_4
+ \lambda_6 \sqrt{\frac{\lambda_1}{\lambda}} > 0 ,
\\
&&
\omega_2
\equiv
 2\lambda_2
 + \lambda_6 \sqrt{\frac{\lambda_2}{\lambda}} > 0 , \quad
2\sqrt{\omega_2 \lambda_1}
+ 2\lambda_3
+ \lambda_4
+ \lambda_5 \sqrt{\frac{\lambda_2}{\lambda}} > 0 ,
\\
&&
\omega_{12}
\equiv
 2\lambda_1
 + (2\lambda_3 + \lambda_4)
   \sqrt{\frac{\lambda_1}{\lambda_2}} > 0 , \quad
2\sqrt{\omega_{12} \lambda}
+ \lambda_5
+ \lambda_6 \sqrt{\frac{\lambda_1}{\lambda_2}} > 0 .
\end{eqnarray}
 Mass eigenstates of two charged scalar fields are given by
a mixing angle $\theta_\pm$ as
\begin{eqnarray}
\begin{pmatrix}
 H_1^+\\
 H_2^+
\end{pmatrix}
&\equiv&
 \begin{pmatrix}
  \cos\theta_\pm & -\sin\theta_\pm\\
  \sin\theta_\pm & \cos\theta_\pm
 \end{pmatrix}
 \begin{pmatrix}
  s_1^+\\
  s_2^+
 \end{pmatrix} , \quad
%
%
\tan{2\theta_\pm}
=
 \frac{ 2 \mu_3^2 }{ m_{s_1^\pm}^2 - m_{s_2^\pm}^2 } ,
\end{eqnarray}
where we used $m_{s_1^\pm}^2 \equiv \mu_1^2 + \lambda_5 v^2/2$
and $m_{s_2^\pm}^2 \equiv \mu_2^2 + \lambda_6 v^2/2$
with $v \equiv \sqrt{2} \langle \phi^0 \rangle = 246\,\GeV$.
 Clearly,
$m_{s_1^\pm}^2 > 0$ and $m_{s_1^\pm}^2 > 0$
are necessary for $\langle s_1^+ \rangle = \langle s_2^+ \rangle = 0$.
 Masses of $H_1^+$ and $H_2^+$ are expressed as
\begin{eqnarray}
m_{H_1^\pm}^2
&=&
 \frac{1}{\,2\,}
 \left\{
  m_{s_2^\pm}^2 + m_{s_1^\pm}^2
  -
  \sqrt
  {
   \left( m_{s_2^\pm}^2 - m_{s_1^\pm}^2 \right)^2
   + 4 \mu_3^4
  }
 \right\} ,
\\
%
%
m_{H_2^\pm}^2
&=&
 \frac{1}{\,2\,}
 \left\{
  m_{s_2^\pm}^2 + m_{s_1^\pm}^2
  +
  \sqrt
  {
   \left( m_{s_2^\pm}^2 - m_{s_1^\pm}^2 \right)^2
   + 4 \mu_3^4
  }
 \right\} ,
\end{eqnarray}
where $H_1^\pm$ is defined as the lighter one.
 It is required to satisfy
$m_{s_1^\pm}^2\, m_{s_2^\pm}^2 - \mu_3^4 > 0$
so that $m_{H_1^\pm}^2 > 0$
at $\langle s_1^+ \rangle = \langle s_2^+ \rangle = 0$.
 The LEP experiment
constrains masses of charged scalar fields
to be greater than $73\text{-}107\,\GeV$ at the 95\% confidence level
(see mass limits for $H^\pm$ from doublet fields
and charged sleptons in ref.~\cite{Nakamura:2010zzi}).

\begin{figure}[t]
\begin{center}
\includegraphics[scale=0.7]{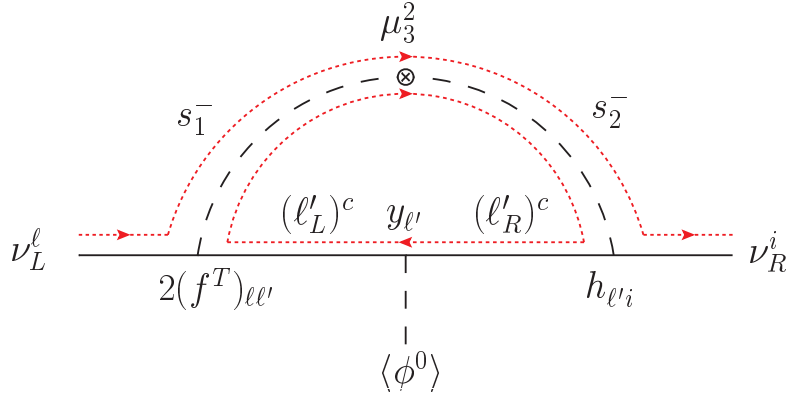}
\vspace*{-5mm}
\caption{
 The one-loop diagram
for Dirac neutrino masses in the \themodel.
 Red dotted arrow shows the flow of lepton number.
}
\label{fig:diagram}
\end{center}
\end{figure}

 In Fig.~\ref{fig:diagram},
the one-loop diagram%
\footnote
{
 The same diagram was used in ref.~\cite{Mohapatra:1987hh}
for the left-right symmetric model,
where Yukawa coupling constant $h_{\ell i}$ in the \themodel\
is replaced by $f_{\ell\pell}$.
 Current neutrino oscillation data
cannot be satisfied in the case.
}
for the Dirac neutrino mass is shown.
 The Dirac neutrino mass matrix $M_{\nu_D}$
for $(M_{\nu_D})_{\ell i}\,\overline{\nu_L^\ell}\,\nu_R^i$
is obtained as follows:
\begin{eqnarray}
(M_{\nu_D})_{\ell i}
=
 C\,
 (f^T)_{\ell\pell}\, m_\pell^{}\, h_{\pell i}^{} , \quad
%
%
C
\equiv
 \frac{\sin{2\theta_\pm}}{16\pi^2} \ln\frac{m_{H_2^\pm}^2}{m_{H_1^\pm}^2} .
\end{eqnarray}
 Needless to say,
$\theta_\pm = 0$ and $\pi/2$
are not acceptable to obtain nonzero neutrino masses. 
 Since we are taking the basis
where $\ell$ and $\nu_R^i$ are mass eigenstates,
the Dirac mass matrix can be expressed as
\begin{eqnarray}
(M_{\nu_D})_{\ell i}
=
 (U_\MNS)_{\ell i}\, m_i ,
\end{eqnarray}
where $m_i$ ($i=1\text{-}3$) are
neutrino mass eigenvalues ($m_i \geq 0$).
 The matrix $U_\MNS$ is
the Maki-Nakagawa-Sakata (MNS) matrix~\cite{Maki:1962mu}
which is expressed in the standard parametrisation as
\begin{eqnarray}
U_\MNS
=
 \begin{pmatrix}
  1 & 0 & 0\\
  0 & c_{23} & s_{23}\\
  0 & -s_{23} & c_{23} 
 \end{pmatrix}
 \begin{pmatrix}
  c_{13} & 0 & s_{13}\, e^{-i\delta}\\
  0 & 1 & 0\\
  -s_{13}\, e^{i\delta} & 0 & c_{13} 
 \end{pmatrix}
 \begin{pmatrix}
  c_{12} & s_{12} & 0\\
  -s_{12} & c_{12} & 0\\
  0 & 0& 1 
 \end{pmatrix} ,
\label{eq:MNS}
\end{eqnarray}
where $c_{ij}$ and $s_{ij}$ stand for
$\cos{\theta_{ij}}$ and $\sin{\theta_{ij}}$,
respectively.
 One of the $m_i$'s is zero in this model
because $\text{Det}(M_{\nu_D}) \propto \text{Det}(f^T)=0$.
 Current neutrino oscillation data%
~\cite{solar,Wendell:2010md,acc,Apollonio:2002gd,Gando:2010aa}
allow two choices;
either $m_1=0$ or $m_3=0$.
 Two squared mass differences $\Delta m^2_{ij} \equiv m_i^2 - m_j^2$
are taken as
$\Delta m^2_{21} = 7.5\times 10^{-5}\,\eV^2$
and $|\Delta m^2_{31}| = 2.3\times 10^{-3}\,\eV^2$.

 For $m_1=0$,
we have the following relations
which were not found in ref.~\cite{Nasri:2001ax}:
\begin{eqnarray}
h_{\mu 1}
&=&
 - \frac{ m_e\, \Ume^\ast }{ m_\mu\, \Uee }\, h_{e 1} ,
\label{eq:hm1-N}\\
%
%
h_{\mu 2}
&=&
 - \frac{ m_e\, \Ume^\ast }{ m_\mu\, \Uee }\, h_{e 2}
 + \frac{ \Utm\, m_2 }{ C\, m_\mu\, f_{\mu\tau} } ,
\label{eq:hm2-N}\\
%
%
h_{\mu 3}
&=&
 - \frac{ m_e\, \Ume^\ast }{ m_\mu\, \Uee }\, h_{e 3}
 + \frac{ \Utt\, m_3 }{ C\, m_\mu\, f_{\mu\tau} } ,
\label{eq:hm3-N}\\
%
%
h_{\tau 1}
&=&
 \frac{ m_e\, \Ute^\ast }{ m_\tau\, \Uee }\, h_{e1} ,
\label{eq:ht1-N}\\
%
%
h_{\tau 2}
+
\frac{ m_\mu\, \Umm }{ m_\tau\, \Utm }\, h_{\mu 2}
&=&
 - \frac{ m_e\, \Uem }{ m_\tau\, \Utm }\, h_{e 2} ,
\label{eq:ht2-N}\\
%
%
h_{\tau 3}
+
\frac{ m_\mu\, \Umt }{ m_\tau\, \Utt }\, h_{\mu 3}
&=&
 - \frac{ m_e\, \Uet }{ m_\tau\, \Utt }\, h_{e 3} ,
\label{eq:ht3-N}\\
%
%
f_{e\mu}
&=&
 \frac{ \Ute^\ast }{ \Uee }\, f_{\mu\tau} ,
\label{eq:fem-N}\\
%
%
f_{e\tau}
&=&
 \frac{ \Ume^\ast }{ \Uee }\, f_{\mu\tau} .
\label{eq:fet-N}
\end{eqnarray}
 When we fix the MNS matrix and neutrino masses,
six elements of $h_{\ell i}$ and two elements of $f_{\ell\pell}$
in the left-hand side of these equations
are given by five variables
($h_{e1}$, $h_{e2}$, $h_{e3}$, $f_{\mu\tau}$, and $C$).
 Rephasing of the massless $\nu_R^1$ makes $h_{e1}$ real.
 Two phase degrees of freedom remain in $h_{\ell i}$.
 However,
eqs.~(\ref{eq:fem-N}) and (\ref{eq:fet-N})
show that the CP violating phase $\delta$ in the MNS matrix
vanishes because $f_{\ell\pell}$ are real.

 We assume for simplicity
the so-called tribimaximal mixing~\cite{Harrison:2002er}
($s_{23}^2 = 1/2, s_{12}^2=1/3, s_{13}^2=0$)
which agrees with neutrino oscillation data well.
 A simple example of the parameter set (the set~A)
which satisfy eqs.~(\ref{eq:hm1-N})-(\ref{eq:fet-N}) is
\begin{eqnarray}
h
&=&
 1.1\times 10^{-2}
 \begin{pmatrix}
  1
  & 1.8\times 10^{-1}
  & 1
\\
  -2.4\times 10^{-3}
  & -9.4\times 10^{-4}
  & 1.0\times 10^{-3}
\\
  1.4\times 10^{-4}
  & -4.0\times 10^{-6}
  & 5.9\times 10^{-5}
 \end{pmatrix} ,
\label{eq:hA}
\\
f
&=&
 1.1\times 10^{-2}
 \begin{pmatrix}
  0
  & 0.5
  & 0.5
\\
  -0.5
  & 0
  & 1
\\
  -0.5
  & -1
  & 0
 \end{pmatrix} ,
\label{eq:fA}
\\
m_{H_1^\pm}^{}
&=&
 150\,\GeV , \quad
m_{H_2^\pm}^{}
=
 200\,\GeV , \quad
\theta_\pm
=
 0.1\,\text{rad} .
\label{eq:mHval}
\end{eqnarray}
 Notice that
some elements of $h_{\ell i}$ (especially, $h_{\tau i}$)
tend to be small because of ratios of charged lepton masses
in eqs.~(\ref{eq:hm1-N})-(\ref{eq:ht3-N})
while all elements of $f_{\ell\pell}$ are in the same order of magnitude.

 For $m_3=0$,
we obtain
\begin{eqnarray}
h_{e1}
&=&
 \frac{ m_\mu\, \Uet^\ast }{ m_e\, \Umt }\, h_{\mu 1}
 + \frac{ \Ute\, m_1 }{ C\, m_e\, f_{e\tau} } ,
\label{eq:he1-I}\\
%
%
h_{e2}
&=&
 \frac{ m_\mu\, \Uet^\ast }{ m_e\, \Umt }\, h_{\mu 2}
 + \frac{ \Utm\, m_2 }{ C\, m_e\, f_{e\tau} } ,
\label{eq:he2-I}\\
%
%
h_{e3}
&=&
 \frac{ m_\mu\, \Uet^\ast }{ m_e\, \Umt }\, h_{\mu 3} ,
\label{eq:he3-I}\\
%
%
h_{\tau 1}
+
\frac{ m_e\, \Uee }{ m_\tau\, \Ute }\, h_{e1}
&=&
 - \frac{ m_\mu\, \Ume }{ m_\tau \Ute }\, h_{\mu 1} ,
\label{eq:ht1-I}\\
%
%
h_{\tau 2}
+
\frac{ m_e\, \Uem }{ m_\tau\, \Utm }\, h_{e 2}
&=&
 - \frac{ m_\mu\, \Umm }{ m_\tau\, \Utm }\, h_{\mu 2} ,
\label{eq:ht2-I}\\
%
%
h_{\tau 3}
&=&
 \frac{ m_\mu\, \Utt }{ m_\tau\, \Umt }\, h_{\mu 3} ,
\label{eq:ht3-I}\\
%
%
f_{e\mu}
&=&
 - \frac{ \Utt }{ \Umt }\, f_{e \tau} ,
\label{eq:fem-I}\\
%
%
f_{\mu\tau}
&=&
 - \frac{ \Uet^\ast }{ \Umt }\, f_{e \tau} .
\label{eq:fmt-I}
\end{eqnarray}
 Notice that eq.~(\ref{eq:ht2-I}) is the same as eq.~(\ref{eq:ht2-N}).
 The phase of $h_{\mu 3}$ is absorbed by the massless $\nu_R^3$
while two phase degrees of freedom remain in $h_{\mu 1}$ and $h_{\mu 2}$.
 Equation~(\ref{eq:fmt-I})
means $\delta = 0$
similarly to the case of $m_1=0$.
 It is worth to mention that
we obtain $s_{23}^2 = 1/2$ and $s_{13}^2=0$
independently of $h_{\ell i}$
if $f_{e\mu}=-f_{e\tau}$ and $f_{\mu\tau}=0$, respectively.
 Such conditions on $f_{\ell\pell}$ might be given
by some discrete symmetry.
 Equations~(\ref{eq:he1-I})-(\ref{eq:fmt-I})
for the tribimaximal mixing
are satisfied by the following example (the set~B)
with values in eq.~(\ref{eq:mHval}):
\begin{eqnarray}
h
&=&
 8.7\times 10^{-3}
 \begin{pmatrix}
  -7.0\times 10^{-1}
  & 1
  & 0
\\
  1
  & 1
  & 1
\\
  6.0\times 10^{-2}
  & 6.0\times 10^{-2}
  & 6.0\times 10^{-2}
 \end{pmatrix} ,
\label{eq:hB}
\\
f
&=&
 8.7\times 10^{-3}
 \begin{pmatrix}
  0
  & 1
  & -1
\\
  -1
  & 0
  & 0
\\
  1
  & 0
  & 0
 \end{pmatrix} .
\label{eq:fB}
\end{eqnarray}

\section{phenomenology}
\label{sec:pheno}

 In this section,
we consider the constraint from the lepton flavor violating~(LFV) decays
of charged leptons
and the prospect for the LHC physics.

\subsection{Lepton flavor violation}
\label{subsec:LFV}

 The most stringent constraint on this model
from the LFV decays of charged leptons
is given by the experimental bound on the branching ratio~(BR) of $\meg$,
$\BR(\meg) < 1.2\times 10^{-11}$~\cite{Brooks:1999pu}.
 The branching ratio in this model
is calculated as
\begin{eqnarray}
\BR(\mu \to e \gamma)
\simeq
 \frac{ \alpha }{ 768 \pi G_F^2 }
 \left\{
  16 f_{e\tau}^2 f_{\mu\tau}^2\!
  \left(
   \frac{ c_\pm^2 }{ m_{H_1^\pm}^2 } + \frac{ s_\pm^2 }{ m_{H_2^\pm}^2 }
  \right)^2\!\!\!
  +
  \left| (h h^\dagger)_{\mu e} \right|^2\!
  \left(
   \frac{ s_\pm^2 }{ m_{H_1^\pm}^2 } + \frac{ c_\pm^2 }{ m_{H_2^\pm}^2 }
  \right)^2
 \right\} ,
\label{eq:BRmeg}
\end{eqnarray}
where $c_\pm \equiv \cos\theta_\pm$ and $s_\pm \equiv \sin\theta_\pm$.
 We ignore fermion masses in the loop integration
and the electron mass in the final state.
 The parameter set~A (eqs.~(\ref{eq:hA})-(\ref{eq:mHval}))
results in $\BR(\meg)=2.9\times 10^{-12}$.
 This means
not only that
the \themodel\ can satisfy the current bound on $\BR(\meg)$
but also that the BR can be in the expected sensitivity
of experiments in the future.
 On the other hand,
the set~B (eqs.~(\ref{eq:mHval}), (\ref{eq:hB}) and (\ref{eq:fB}))
satisfies the experimental bound
with a much smaller value $\BR(\meg)=7.5\times 10^{-15}$.
 This is because $f_{\mu\tau}$ and $h_{e3}$ for $m_3=0$ are
proportional to small $s_{13}$.
 Although $\BR(\tau\to\mu\gamma)\sim 10^{-13}$ for the set~B
is much larger than $\BR(\meg)$,
it is also far from experimental sensitivity.

 In the \themodel\
the coupling constants $f_{\ell\pell}$ and $h_{\ell i}$
can be ${\mathcal O}(10^{-2})$.
 Then
the box diagram contributions to $\mu \to \bar{e}ee$,
which are proportional to the eighth power of these coupling constants,
can be smaller than the current experimental upper bound
although it becomes crucial in models
where some of coupling constants are ${\mathcal O}(1)$~\cite{Aoki:2011zg}.

\subsection{Prospects at the LHC}
\label{subsec:LHC}

 The charged scalar boson $H_1^\pm$
is expected to be produced at the LHC
if it is light.
 The production cross section
via $q\bar{q} \to \gamma^\ast, Z^\ast \to H_1^+ H_1^-$
with $\sqrt{s} = 14\,\TeV$
is 23\,fb for $m_{H_1^\pm} = 150\,\GeV$ for example.
 The partial decay widths of
$H_1^- \to \ell\nu$ ($\ell = e,\mu,\tau$) are given by
\begin{eqnarray}
\Gamma(H_1^- \to \ell \nu)
\equiv
 \sum_{\pell}
 \Gamma(H_1^- \to \ell \nu_{\pell}^{})
\simeq
 \frac{m_{H_1^\pm}}{16\pi}
 \left(
  4 c_\pm^2 \sum_\pell |f_{\ell\pell}|^2
  + s_\pm^2 \sum_i |h_{\ell i}|^2
 \right) ,
\end{eqnarray}
where fermion masses are neglected.

 If $H_1^\pm$ is made dominantly from $s_1^\pm$,
its decay is determined by $f_{\ell\pell}$.
 By using eqs.~(\ref{eq:fem-N}) and (\ref{eq:fet-N}) for $m_1=0$
and the tribimaximal mixing,
we obtain
\begin{eqnarray}
\BR(H_1^- \to e \nu)
: \BR(H_1^- \to \mu \nu)
: \BR(H_1^- \to \tau \nu)
=
 2:5:5 .
\end{eqnarray}
 The set~A (eqs.~(\ref{eq:hA})-(\ref{eq:mHval})) gives
approximately the same result.
 Equations~(\ref{eq:fem-I}) and (\ref{eq:fmt-I}) for $m_3=0$
and the tribimaximal mixing
give
\begin{eqnarray}
\BR(H_1^- \to e \nu)
: \BR(H_1^- \to \mu \nu)
: \BR(H_1^- \to \tau \nu)
=
 2:1:1 .
\end{eqnarray}
 The set~B (eqs.~(\ref{eq:hB}), (\ref{eq:fB}), and (\ref{eq:mHval}))
gives the same result in a good approximation.
 These results are robust
because the matrix structure of $f_{\ell\pell}^{}$
is restricted very well.
 Therefore,
if $H_1^\pm \simeq s_1^\pm$,
this model predicts
$\BR(H_1^- \to \tau \nu)/\BR(H_1^- \to \mu \nu) \simeq 1$.
 On the other hand,
if $H_1^\pm$ is made dominantly from $s_2^\pm$,
its partial decay widths are controlled by $h_{\ell i}$.
 For $m_1 = 0$,
eqs.~(\ref{eq:hm1-N}) and (\ref{eq:ht1-N})
show $h_{\tau 1} \sim h_{\mu 1} m_\mu/m_\tau$.
 Furthermore,
we have $h_{\tau 3} \sim h_{\mu 3} m_\mu/m_\tau$
with eq.~(\ref{eq:ht3-N}) for $\theta_{13}=0$.
 For $m_3 = 0$,
eq.~(\ref{eq:ht3-I}) also means
$h_{\tau 3} \sim h_{\mu 3} m_\mu/m_\tau$.
 Thus,
it seems reasonable to expect
$h_{\tau i} \sim h_{\mu i} m_\mu/m_\tau$ ($i=1\text{-}3$).
 Then
we have
\begin{eqnarray}
\frac{ \BR(H_1^- \to \tau \nu) }{ \BR(H_1^- \to \mu \nu) }
\sim
 \frac{m_\mu^2}{m_\tau^2}
\sim
 10^{-2} .
\end{eqnarray}
 As the result,
this model is likely to give
$\BR(H_1^- \to \tau \nu)/\BR(H_1^- \to \mu \nu) \lesssim 1$
due to the discussion above.

 If $H_2^-$ is also light and
the production cross section is significant,
the decays into $\ell \nu$ ($\ell = e,\mu,\tau$)
smear the relation discussed above to some extent.
 Otherwise
we can test the model at the LHC as well as the ILC
by measuring the above characteristic pattern
of the decay branching ratios.

 The partial decay width for $h^0 \to \gamma\gamma$
of the SM Higgs boson $h^0$,
which is caused at the one-loop level in the SM,
is affected by additional one-loop diagrams
with $H_1^+$ and $H_2^+$.
 The contributions to the SM prediction
depend on $\lambda_5$ and $\lambda_6$
as well as Higgs masses.
 When coupling constant for $h^0H_i^+H_i^-$~($i=1$ or $2$)
is positive (negative),
the additional loop effect from $H_i^+$ gives
a destructive (constructive) contribution to the SM prediction.
 These contributions can amount
to ${\mathcal O}(10)\,\%$ deviations~\cite{Kanemura:2000bq}.
 If such the effect is detected at the LHC
when the light SM Higgs boson is discovered,
it can be an important indirect signature of the charged singlet scalar bosons.

\section{discussions}
\label{sec:disc}

 Possible extensions of the \themodel\ are discussed in this section.
 First,
we try to introduce dark matter candidates
which do not exist in the model.
 Next,
we consider the case with lepton number violation.
 Even if lepton number is not conserved
and Majorana mass terms for $\nu_R^i$ are allowed,
the mechanism to suppress the Dirac mass term is fruitful.

\subsection{Introducing dark matter candidates}
\label{subsec:DM}

 A possibility to accommodate dark matter candidates
would be to impose an unbroken $Z_2$ symmetry
(we call it $Z_2^\prime$)
to this model in addition to the softly-broken $Z_2$ symmetry
such that all particles in the loop are $Z_2^\prime$-odd%
\footnote{
 Instead of the $Z_2^\prime$ symmetry,
lepton number can be used when it is conserved.
 For example,
a fermion (boson) with a lepton number 2 (1) could be stable.
 From this point of view,
lepton number conservation seems fit well
for introducing dark matter candidates.
}.
 Since the SM charged leptons in the loop in Fig.~\ref{fig:diagram}
cannot be $Z_2^\prime$-odd,
they must be replaced by newly introduced
$Z_2^\prime$-odd fermions
which can be understood as the fourth generation leptons.
 The $Z_2^\prime$-odd Dirac neutrino
could be the lightest $Z_2^\prime$-odd particle
which is stable.
 However,
it cannot be identified as the dark matter
because the spin-independent scattering cross section on a nucleon
is too large to satisfy current data of direct searches~\cite{directDM}
due to the diagram mediated by the $Z$ boson.
 Therefore,
such a minimal extended model is excluded.

  We may consider the other model
by taking different scalar particle contents,
where the dark matter candidate enters
and Dirac neutrino masses are induced radiatively.
 Such a model can be found in ref.~\cite{Gu:2007ug}
where the following exact $Z_2^\prime$-odd particles are introduced:
an $SU(2)_L$-doublet scalar field ($\Phi_2 = (\phi_2^+, \phi_2^0)^T$)
and a real neutral singlet scalar ($s^0$)
as well as a neutral singlet Dirac fermion ($N$).
 In ref.~\cite{Gu:2007ug},
a real scalar dark matter ($\text{Re}(\phi_2^0)$ or $s^0$)
and the so-called Dirac leptogenesis~\cite{Dick:1999je}
via $N$ decay are considered.
 If the Dirac fermion $N$ is the lightest $Z_2^\prime$-odd particle,
the dark matter is different from its anti-particle
in contrast with the dark matter of the Majorana particle.
 It could be compatible with
the asymmetric dark matter scenario~\cite{Kaplan:2009ag}.
 The Dirac leptogenesis would be also achieved by the decay of $\phi_2^0$.

 Another simple possibility
is the $R$-parity-conserving SUSY extension
where the lightest supersymmetric particle becomes a candidate
for dark matter.
 The detailed study will be presented elsewhere.

\subsection{Baryogenesis}
\label{subsec:baryon}

 There seem to be two possible extensions of this model
in order to realise baryogenesis,
although the detailed analysis on them
is beyond the scope of this letter.
 One is the electroweak baryogenesis~\cite{Kuzmin:1985mm}.
 The scalar sector should be extended
in order to have CP-violating phases
and also to achieve strong first order phase transition
at the electroweak symmetry breaking.
 The other is an application of the Dirac leptogenesis~\cite{Dick:1999je}.
 It is known that
the leptogenesis is possible without LNV\@.
 The number to be converted to the baryon asymmetry
is free from the $\nu_R^i$ number
because the sphaleron does not act on
the gauge singlet fields $\nu_R^i$.

\subsection{Lepton number violation}
\label{subsec:LNV}

 The mechanism to induce the Dirac mass terms for neutrinos
can be applied also to the lepton number violating case,
in which $\nu_R^i$ have Majorana masses as
$M_i \overline{(\nu_R^i)^c} \nu_R^i$.
 Then,
the type-I seesaw mechanism is realized
at the two-loop level
via the one-loop induced Dirac masses.
 This model was studied in refs.~\cite{Nasri:2001ax,Nasri:2001nb}%
\footnote
{
 A one-particle-irreducible two-loop diagram
also exists for the Majorana masses of $\nu_L$,
which seems to be overlooked in ref.~\cite{Nasri:2001ax}.
}.
 By this loop suppression mechanism,
$M_i$ are much lighter than
those in the tree-level seesaw mechanism.
 Consequently,
$M_i$ can be at the TeV scale
without excessive fine tuning on coupling constants.
 Such TeV scale Majorana neutrinos
could be tested at the LHC as well as the ILC\@.

 In the two-loop seesaw model,
we can remove the soft-breaking term of the $Z_2$ symmetry.
 Then
the Majorana masses for $\nu_L$ are generated at the three-loop level
and the lightest $\nu_R$ becomes a dark matter candidate.
 This model coincides
with the model proposed by Krauss, Nasri and Trodden~\cite{Krauss:2002px}.

\section{conclusions}
\label{sec:concl}

 We have investigated a simple model (\themodel)
with the mechanism for radiative generation of Dirac neutrino masses
without introducing lepton number violation.
 In the \themodel,
the Yukawa interaction
$\overline{L}\,\tilde{\Phi}\,\nu_R$
is absent at the tree level
because of the softly-broken $Z_2$ symmetry,
so that it is induced at the one-loop level
by the soft-breaking in the mixing between
$s_1^\pm$ and $s_2^\pm$.
 Tiny neutrino masses are generated
from the TeV scale dynamics.
 We have found the model can be compatible with
the current neutrino oscillation data
as well as LFV search results (especially for $\meg$).
 There is no CP-violating phase in the MNS matrix in this model.
 It is possible that $\BR(\meg)$
becomes large enough to be discovered by experiments
in near future.
 The \themodel\ is likely to give
$\BR(H_1^- \to \tau \nu)/\BR(H_1^- \to \mu \nu) \lesssim 1$.
 Characteristic features of $H_1^\pm$
are expected to be tested at the LHC and the ILC\@.
 We also have discussed several possible extensions
of the model, which implement dark matter candidates,
mechanism for baryogenesis,
and the radiative type-I seesawlike scenario
by using one-loop suppressed Dirac masses.

\begin{acknowledgments}
 The work of S.K.\ was supported by
Grant-in-Aid for Scientific Research (A)
No.\ 22244031.
 The work of H.S.\ was supported
by the Sasakawa Scientific Research Grant
from the Japan Science Society
and Grant-in-Aid for Young Scientists (B)
No.\ 23740210.
\end{acknowledgments}


\end{document}